\documentclass{emulateapj}
\usepackage{epsfig}
\usepackage[usenames,dvipsnames]{color}
\usepackage{array}
\usepackage{graphicx}
\usepackage{dcolumn}
\usepackage{bm}
\usepackage{epsfig}
\usepackage{amssymb,amsmath}
\usepackage{natbib}
\usepackage{url}
\usepackage{rotating}
\usepackage{subfigure}
\begin{document}

\title{Lukewarm dark matter: Bose condensation of ultralight particles}
\author{Andrew P. Lundgren \altaffilmark{1}, Mihai Bondarescu \altaffilmark{2}, Ruxandra Bondarescu \altaffilmark{3}, Jayashree Balakrishna \altaffilmark{4}}
\altaffiltext{1}{Syracuse University, Syracuse, NY}
\altaffiltext{2}{University of Mississippi, Oxford, MS}
\altaffiltext{3}{Pennsylvania State University, State College, PA}
\altaffiltext{4}{Harris Stowe State University, St Louis, MO}

\begin{abstract}
We discuss the thermal evolution and Bose-Einstein condensation of ultra-light dark matter particles at finite, 
realistic cosmological temperatures. We find that if these particles decouple from regular matter before Standard 
model particles annihilate,  their temperature will be about 0.9 K. This temperature is substantially lower than 
the temperature of CMB neutrinos and thus Big Bang Nucleosynthesis remains unaffected. 
In addition the temperature is consistent with WMAP 7-year+BAO+H0 observations without fine-tuning. 
We focus on particles of mass of $m\sim 10^{-23}$ eV, which have Compton wavelengths of galactic scales.
Agglomerations of these particles can form stable halos and naturally prohibit small scale structure.
They avoid over-abundance of dwarf galaxies and may be favored by observations of dark matter distributions.  
We present numerical as well as approximate analytical solutions of the Friedmann-Klein-Gordon equations 
and study the cosmological evolution of this scalar field dark matter from the early universe to the 
era of matter domination.  Today, the particles in the ground state mimic presureless matter, while 
the excited state particles are radiation like. 
\end{abstract}
\keywords{(cosmology:) theory, (cosmology:) dark matter, (cosmology:) early universe}
\maketitle
\section{Introduction}
Most matter in the universe is non-luminous. The observed flatness of the galactic rotation
curves indicates the presence of dark matter halos around galaxies. Observations of the cosmic microwave background anisotropies \citep{WMAP} combined with 
large-scale structure and type Ia supernova luminosity data  \citep{Reiss, Per1999} constrain cosmological parameters finding that visible matter contributes 
only about 4\% of the energy density of the universe, as opposed to 22\% being dark matter and 74\% dark energy. More recently, a clear separation between 
the center of baryonic matter and the total center of mass was observed in the Bullet cluster \citep{BulletCluster} and later in other galaxy 
cluster collisions \citep{OtherCluster}. These observations reinforce the claim that dark matter is indeed composed of weakly interacting particles and 
is not a modification of gravity. 

In the past few decades numerous dark matter candidates have been suggested including WIMPs, axions, 
and various spin zero bosons \citep{Kamin07, Ber2005}.   Fundamental spin zero particles 
represented by scalar fields play an important role in particle physics models \citep{PQ1977,Torres00}. These particles could form gravitationally stable
structures such as boson stars, soliton stars, and galactic halos 
through some type of Jeans instability mechanism \citep{Jeans1,Jeans3,Jeans2}. The stability and gravitational wave signatures of compact scalar stars 
have been studied numerically \citep{K1985,edwm, bala2006, bala2008}.

Bosonic halos in which scalar particles Bose-condense and form gravitationally stable structures
are supported against collapse by Heisenberg's uncertainty principle like boson stars.
Structure formation on scales smaller than the spreading of an individual boson (the Compton wavelength of one particle) is forbidden by 
quantum mechanics \citep{Hu,Varun,Lee08}. Halos formed from ultralight scalars with Compton wavelength of galactic scales thus do not 
lead to over-abundance of dwarf galaxies unlike cold dark matter simulations with heavier bosons \citep{overabundance1, Matos01,miguel02,overabundance2}.
   
  Scalar field halos have been fit to rotation curves of spiral galaxies \citep{Franz1,GL,Matos00,arby1,Boehmer07}. 
By using the mass as a free parameter to fit rotation 
curves \cite{arby1} have obtained $m = (0.4-1.6) \times 10^{-23}$ eV for non-interacting ultra-light bosons.   
In a followup paper, \cite{arby} performed a non-thermal analysis of ultralight bosonic halos.

   \cite{urena-lopez} pointed out that scalars field particles can Bose-condense at finite temperatures resurrecting previous work on relativistic Bose-Einstein condensation by  \cite{ParkerZhang1, ParkerZhang2}. A condensate is considered relativistic
   when  the temperature of the condensate is significantly larger than the mass of one boson. \cite{ParkerZhang2} discuss inflationary expansion driven by a relativistic Bose-Einstein condensate that then evolves into a radiation dominated universe. 
        
 In this paper we perform a thermal analysis of the post-inflationary cosmological behavior of scalar field dark matter formed from ultra-light bosons. 
 We use the quantum field theory formalism of \cite{PF} and extend the analysis 
 that \cite{Hu1980s} used for a description of finite temperature effects in the early universe. 
 The bosons are described by a complex scalar field to provide a conserved charge.  

We assume the scalar particles decouple after inflation in the early universe, after which
the field has a simple quadratic potential with no interactions. There are no constraints on this field 
from particle physics or precision tests of gravity, as there would be for an interacting light scalar field. 
Ultralight particles ($m \sim 10^{-23}$ eV) form a pure ground state Bose-Einstein condensate with a high critical 
temperature behaving like cold dark matter today. Particles in excited states behave like radiation today and hence 
contribute to the amount of hot dark matter in the universe, which is constrained by cosmological observations. 
Our model assumes that the interaction of the scalar field with normal matter turns off in the early universe, at 
a temperature where most Standard Model particles have not yet annihilated, which yields a scalar field temperature 
$T_\Phi = 0.9$ K with no fine-tuning. A reasonable choice of $T_\Phi$ strongly consistent
with cosmological observations, is $T_\Phi \lesssim 1.5$ K. Since these estimates come from states that behave 
like radiation, they are independent of particle mass.
  
 In \S \ref{sec2} we review relativistic Bose-condensation at high temperatures and discuss the temperature at which these scalars could decouple.  \S \ref{cosevol}   follows the cosmological evolution of the field. We use units where $\hbar=c = 1$.  
   
\section{Bose-Einstein condensation at finite temperature}
\label{sec2}
We consider a system of ultralight ($m \sim 10^{-23}$ eV) relativistic bosons represented by complex scalar fields. The condition for a relativistic condensate is that the temperature of the condensate $T\gg m$, which is certainly true up to the present day \citep{urena-lopez}.

\footnotetext[1]{Recently, \cite{Sikivie} showed that dark matter axions can form a BEC as well.}
 In the case of a complex field, there is a conserved charge, which is required for traditional Bose-Einstein condensation (BEC)$^1$.  The charge density is defined as the excess of particles $n$ over anti-particles $\bar{n}$:
\begin{equation}
q = n - \bar{n}.
\end{equation}

For the excited states, this charge density is \citep{Mukhanov}
\begin{equation}
q_{\rm ex} = g \frac{\mu T^2}{3},
\end{equation}
where $g$ is the number of degrees of freedom of the system and the chemical potential 
$\mu \le m$.  The maximum $q_{\rm ex} = m T^2/3$ occurs for $\mu = m$. For ultra-light bosons, the excess 
of bosons over anti-bosons in the excited states is small compared to the number density. 
Any charge added to the system when $\mu = m$ must condense to the ground state. If`$q$ is large, 
the ground state is populated almost exclusively by particles \citep{Mukhanov}.

The critical temperature below which condensation occurs is found in terms of the charge density 
of the dark matter particles \citep{urena-lopez, Mukhanov}
\begin{equation}
T < T_c = \sqrt{\frac{3 q}{m}} ~.
\label{TcEq}
\end{equation}
When $T<T_c$ the majority of the bosons will condense to the ground state. In the ground state the particles will behave like non-relativistic matter, while the particles in excited states will remain highly relativistic.  

 Assuming that BEC occurs and that most particles are in the ground state, a first approximation to the total dark matter density is
\begin{equation}
\rho_{DM} \approx (n + \bar{n}) m.
\end{equation}

The density of dark matter today $\rho_{DM}^{\rm 0}$ is
\begin{equation}
\rho_{DM}^{\rm 0} \approx 23 \% \rho_c, 
\end{equation}
where $\rho_c \approx 4.19 \times 10^{-11} eV^4$. Since $n\gg\bar{n}$ 
\begin{eqnarray}
 n \approx \frac{\rho_{DM}}{m} \approx 10^{12} eV^3
\end{eqnarray}
and
\begin{equation}
T_c \approx  1.7 \times 10^{17} eV \sim 10^{21} \; \rm K,
\end{equation}
which corresponds to a very pure condensate today. Note that the required charge density is very high implying the necessity of a mechanism that would produce such a large asymmetry of scalar particles over anti-particles.

The matter in the excited states is relativistic and contributes to the density of hot dark matter $\rho_{HDM}$, commonly 
parameterized by the effective number of neutrino species $N_{eff}$. The nominal value due to the three known neutrino 
species plus small cosmological corrections is $N_{eff}=3.04$. The scalar field temperature $T_{\Phi}$ can be determined from
\begin{equation}
\rho_{HDM} = \left( N_{eff} - 3.04 \right) \frac{7 \pi^2}{120} T_{\nu}^4 = \frac{\pi^2}{15} T_{\Phi}^4 ~,
\end{equation}
with $T_{\nu}=1.95$ K the neutrino temperature today. The value given by the WMAP 7-year results combined with 
measurements of the Baryon Acoustic Oscillation and Hubble parameter \citep{Komatsu2010} is 
$N_{eff}=4.34^{+0.86}_{-0.88}$ (68\% confidence level). 
The 68\% confidence level on $N_{eff}$ corresponds to $T_\Phi$ between $1.51$ K and $2.25$ K. 

Big Bang Nucleosynthesis measurements constrain relic abundances of deuterium and 
$^4He$ putting a 2-sigma bound of $N_{eff}<3.5$ (or $N_{eff}<3.3$), which depends on the uncertainty in 
the $^4He$ abundance considered \citep{BBN}. This bound corresponds to $T_\Phi<1.5$~K ($T_\Phi<1.35$~K).
Larger values of $N_{eff}$ allow a greater contribution of unknown particles like our scalar field to $\rho_{HDM}$.  
This could either allow $T_\Phi$ to be larger or allow other particles to contribute to $\rho_{HDM}$.
 
We now show that $T_\Phi$ can be less than the observational upper limits and substantially less than $T_\nu$, without fine-tuning. 
 Our model assumes that the scalar field interactions with normal matter turned off completely in the very early universe.  
Subsequently, the heavier Standard Model particles annihilated, dumping their energy into the photons, neutrinos, electrons, 
and positrons that remained, increasing their temperature relative to that of the scalar field. The photons and 
leptons have about $g_\ell=10.75$ degrees of freedom, the complex scalar field has $2$, and  when all the Standard Model particles 
are present $g_{*} \approx 100$.  Following a standard textbook \citep{Mukhanov}, entropy conservation gives a relation for the temperatures
\begin{equation}
g_\ell T_\nu^3 + 2 T_\Phi^3 = g_{*} T_\Phi^3 ~.
\end{equation}
This gives $T_\Phi \approx 0.9 K$, well within our constraints.
   
\section{Cosmological Evolution}
\label{cosevol}
The density evolution of the universe must closely follow the standard $\Lambda$CDM model 
at times later than nucleosynthesis, at $a \approx 10^{-10}$.  During radiation domination, our 
scalar field must be a subdominant contribution to the density of the universe, and during matter 
domination it must be a replacement for dark matter.  As shown below, the excited states are a 
subdominant contribution to the density.  The macroscopically-occupied ground state has 
$\rho \propto a^{-6}$ at early times and must be constrained to be less dense than 
the density of radiation for at least all times after nucleosynthesis.

We first derive equations in a general background, as well as expressions for density and pressure, 
then specialize to the epochs of radiation and matter domination.  We also present numerical solutions to the equations 
and show that the initial conditions can be adjusted to satisfy the cosmological constraints.

\subsection{Evolution Equations}

The line element in an expanding universe can be written as the Friedmann-Robertson-Walker (FRW) metric
\begin{equation}
ds^2 = - dt^2 + a(t)^2 \left(dx^2 + dy^2 + dz^2 \right).
\end{equation}
The Klein-Gordon equation 
\begin{eqnarray}
\Box \Phi - m^2 \Phi = 0 \nonumber
\end{eqnarray}
is derived from a Lagrangian density of the form
\begin{equation}
{\cal L} = \frac{1}{2} \sqrt{-g} (g^{\mu \nu} \partial_\mu \Phi \partial_\nu \Phi  - m^2 \Phi).
\end{equation}

The density and pressure can be defined in the usual way
\begin{eqnarray}
\label{prho}
\rho &=& \frac{1}{2} \left( \partial_t \Phi^\dagger \partial_t \Phi +  \partial_j \Phi^\dagger \partial^j \Phi + m^2 \Phi \Phi^\dagger \right), \\ \nonumber
p &=&  \frac{1}{2} \left(\partial_t \Phi^\dagger \partial_t \Phi +  \partial_j \Phi^\dagger \partial^j \Phi -  m^2 \Phi \Phi^\dagger \right),
\end{eqnarray}
where Greek indices vary between 1 and 4 and the index $j$ varies between 1 and 3.

Following \cite{Hu1980s}, we perform a series of variable transformations to expose the conformal properties of the scalar field equation.  We introduce a conformal time coordinate defined by $dt = a d\tau$.  We also make the substitution $\Psi = a \Phi$.  The metric is conformally static
\begin{equation}
ds^2 = a(\tau)^2 \left( -d\tau^2 + dx^2 + dy^2 + dz^2 \right) ~.
\end{equation}
We can now rewrite the Klein-Gordon equation using the flat space operator  $\widetilde{\Box} \equiv -\partial^2_\tau + \partial^2_x + \partial^2_y + \partial^2_z$ and remembering that the d'Alembertian $\Box = (\sqrt{-g})^{-1} \partial_\mu (\sqrt{-g} g^{\mu \nu} \partial_\nu)$

\begin{equation}
 \frac{1}{a^3} \widetilde{\Box} \Psi + \frac{a''}{a^4} \Psi - \frac{m^2}{a} \Psi = 0 ~,
 \label{KG1}
\end{equation}
where $'$ denotes the derivative with respect to $\tau$.
The last two terms in Eq.\ (\ref{KG1}) break conformal invariance. 
The invariance could be restored, as in \cite{Hu1980s}, by setting the mass to zero and adding a term 
proportional to the four-dimensional Ricci scalar.  However, in this case the field would no longer be minimally 
coupled.  We choose to treat the terms that break conformal invariance as a perturbation.

The solution to the scalar field equation can be decomposed into modes \citep{PF}
\begin{equation}
 \Psi({\bf x}, \tau)= \int d^3 {\vec k}  A_{\vec k} \psi_{k}(\tau) e^{i \vec k \cdot \vec x} +{\rm H.c.},
\end{equation}
where H.c. denotes the Hermitian conjugate and $\psi_k$ satisfies
\begin{equation}
\frac{d^2 \psi_k}{d\tau^2} + \left[k^2 - \frac{a''}{a}  + a^2 m^2\right] \psi_k =0  ~.
\label{scalar} 
\end{equation}
The conserved current in mode $k$ can be written as
 \begin{equation}
 J_{0k} = \frac{1}{i} \left(\psi_{k} \partial_\tau \psi^\star_{k} - \psi^\star_{k} \partial_{\tau} \psi_k\right).
 \end{equation}
 The canonical commutation relation of the field $\Phi$ and its conjugate momentum $\Pi$ leads to the usual commutation relations
  \begin{equation}
 [A_{\vec k}, A_{\vec k'}] = 0, \; \;\; [A_{\vec k}, A^\dagger_{\vec k'}] = \delta(\vec k, \vec k'),
 \label{com}
 \end{equation}
when the conserved current is chosen to be $J_{0k} = 1$ for particles and $J_{0k} = -1$ for anti-particles.   The operator $A_{\vec k}$ corresponds to physical particles and the number density of particles is defined to be $n = <A^\dagger_{\vec k} A_{\vec k}>$ \citep{PF}. 

The commutation relations are automatically satisfied if we take $\psi_k$ of the form
 \begin{equation}
 \psi_{\vec k}(\tau) = \frac{1}{\sqrt{2 \omega_{k}}} e^{-i \int^\tau \omega_{k} d\tau'}.
 \label{psibreak}
 \end{equation}
  Each mode is now characterized  by its eigenfunction $\omega_k$ given by
 \begin{eqnarray}
  \label{fullw}
 - \frac{1}{2} (a^2 H)^2 \omega_k \frac{d^2\omega_k}{da^2} + \frac{3}{4} (a^2 H)^2 \left(\frac{d\omega_k}{da}\right)^2 
 \\ \nonumber - \omega_k^2 a H \frac{d (a^2 H)}{da} + \frac{\omega_k}{2} (a^2 H) \frac{d (a^2 H)}{da} \frac{d \omega_k}{da} - \omega_k^4
 \\ \nonumber + (k^2 + a^2 m^2) \omega_k^2 = 0. \nonumber
 \end{eqnarray}
 
To concomitantly solve the Friedmann equation 
\begin{equation}
H^2 = \frac{8 \pi G}{3} \rho
\end{equation}
we approximate the Hubble parameter in different epochs power laws of the form $H= a'/a^2 = H_0 a^{-n}$. 
Here $\rho = \rho_{\rm rad} + \rho_{\Lambda} + \rho_{m}$, the radiation energy density $\rho_{\rm rad} \propto a^{-4}$, the matter density $\rho_m \approx \rho_{\Phi} \propto a^{-3}$ is dominated by dark matter and the dark energy term $\rho_\Lambda = $ constant in a $\Lambda$CDM model. Thus, the exponent is $n=2$  during the radiation domination era and $n=3/2$ during the matter domination era. 
 \begin{figure}
 \begin{center}
 \subfigure[]
 {
 \epsfxsize = 150pt
 \epsfbox{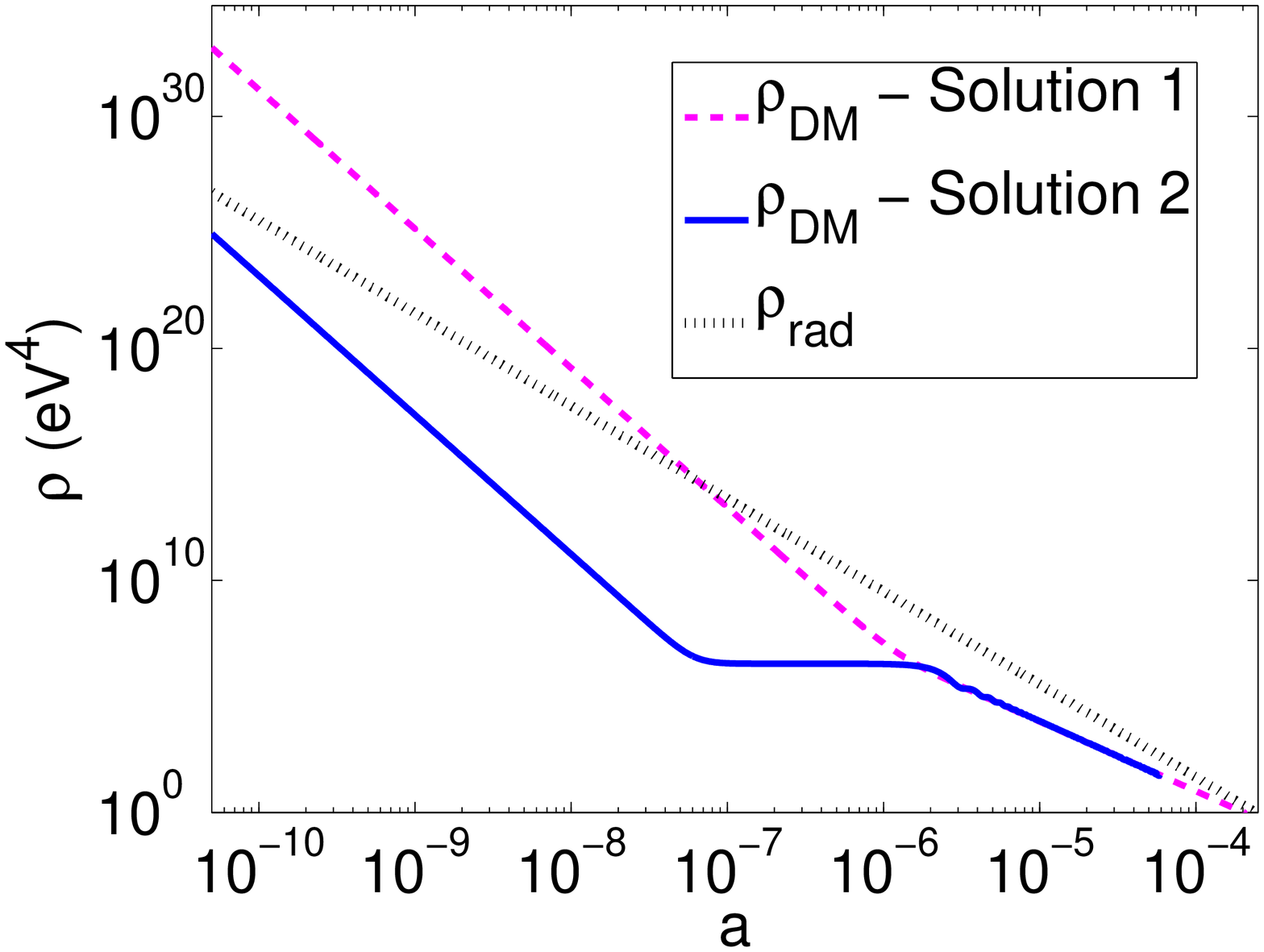}
 }
 \subfigure[]
 {
  \epsfxsize = 150pt
 \epsfbox{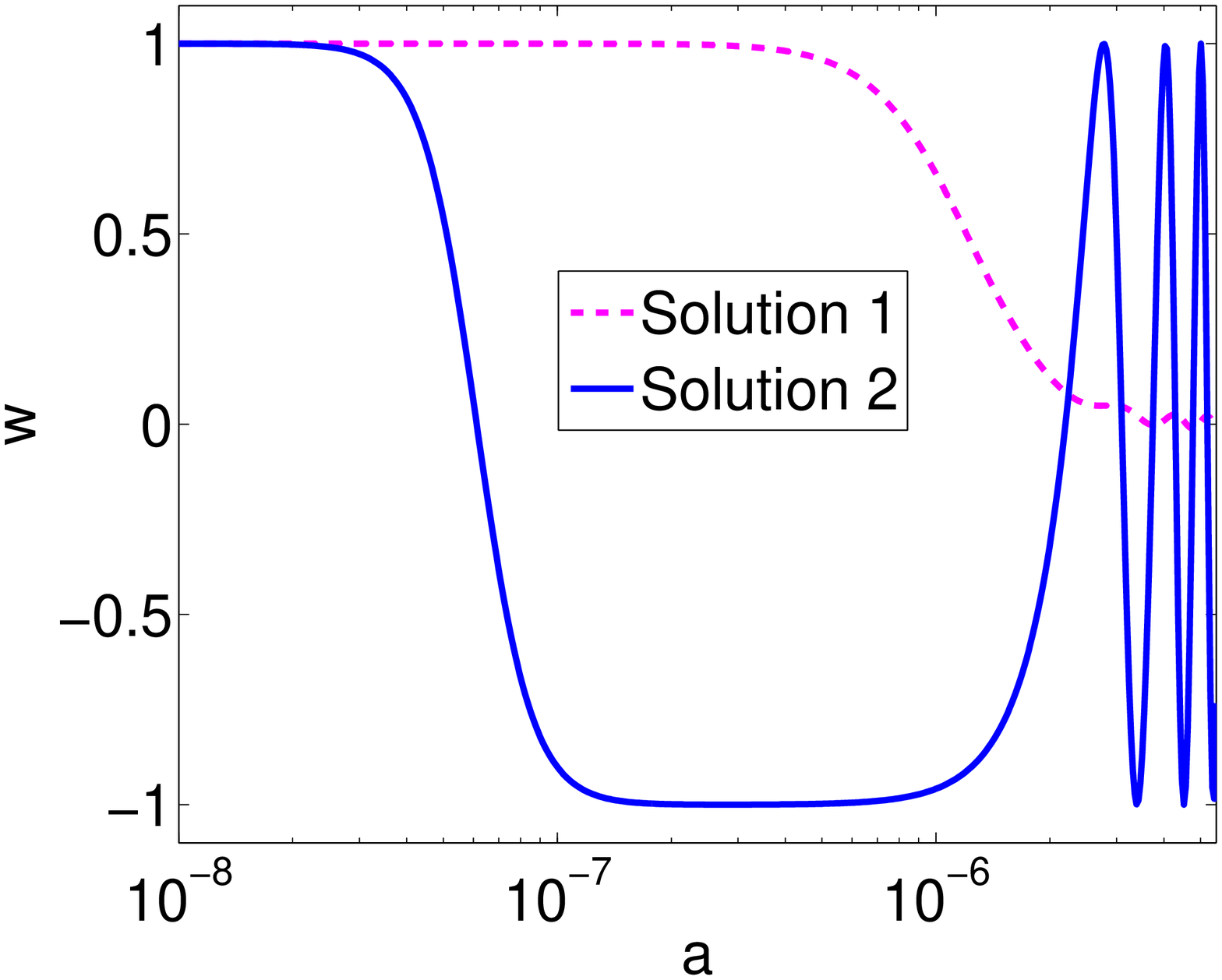}
 }
 \end{center}
 \caption{ The dark matter density $\rho_{DM}$ (along with $\rho_{\rm rad}$) and $w = p_{DM}/\rho_{DM}$ are displayed as a function of scale factor $a$.}
 \label{fig1}
 \end{figure}
 \subsection{Radiation domination}
 In the radiation domination regime ($n=2$), the scalar field Eq.\ (\ref{scalar}) reduces to the flat space wave equation with an effective mass that varies with the scale factor
 
 \begin{equation}
 \label{radpsi}
\frac{d^2 \psi_k}{d\tau^2} + \left(k^2 + a^2 m^2\right) \psi_k =0  ~. 
\end{equation}
 The mass term is the only perturbation from conformal invariance (Since $a'' = 0$.)  
 We extend the analysis of  \cite{Hu1980s} to determine the average density in excited states.   
The density in state $k$ is
 \begin{equation}
 \rho_k = \frac{1}{2} \left[ \frac{H^2}{2 \omega_k a^2} + \frac{(\omega_k')^2}{8 a^4 \omega_k^3} + \frac{H \omega_k'}{2 a^3 \omega_k^2} 
 + \frac{\omega_k}{a^4} \right].
\label{rhok}
 \end{equation}
Modes with $k \gg a m$ are effectively massless ($\omega_k \approx \sqrt{k^2 + a^2 m^2}$).
The total density in the excited states is
 \begin{eqnarray}
 \rho_{\rm ex} &=& \frac{1}{\pi^2} \int_0^\infty k^2 dk \frac{\rho_k}{\exp[\omega_k/(T a)] - 1} \\ \nonumber
   &=& \frac{T^4 \pi^2}{15} + \frac{T^2 H^2}{12} - \frac{m^2}{12} \frac{T^2}{a^2} + ...,
 \end{eqnarray}
where the $\omega_k'$ terms in Eq.\ (\ref{rhok}) contribute to higher order.
 When $T \gg H$, the $T^4$ term dominates the density and the excited states behave like radiation.
 
  For the ground state ($k=0$), Eq. (\ref{fullw}) can be rewritten as
 \begin{equation}
- \frac{y}{2} \frac{d^2 y}{dx^2} + \frac{3}{4} \left(\frac{dy}{dx}\right)^2 - y^4 + x^2 y^2 = 0,
\label{yeq}
 \end{equation}
 where  $y$ and $x$ are dimensionless variables defined by 
  \begin{equation}
\omega_0 =  \sqrt{H_{0r} m} y  , \;\; a = \sqrt{\frac{H_{0r}}{m}} x,
  \end{equation}
  where $H = H_{0r} a^{-2}$ with $H_{0r} \approx 1.4 \times 10^{-35}$ eV.
 Note that $x=1$ ($a \approx 10^{-6}$), which corresponds to $H=m$, is the transition to matter-like behavior for these particles. When $x\ll1$ (or $H\gg m$), we can neglect the  $x^2 y^2$ term. Now Eq.\ (\ref{yeq}) 
 has an exact solution of
 \begin{equation}
 y(x) = \frac{C_0}{1 + C_0^2 (x-x_0)^2},
\label{eq27}
 \end{equation}
 where $C_0$ and $x_0$ are constants.  When $x_0 > 1$, $y$ is approximately constant. When $x_0 \lesssim 1$, the typical behavior is that  $y$ for small $x$, peaks at   $x = x_0$ with a height  $C_0$, and then falls off as $ C_0^{-1} x^{-2}$ (the higher $C_0$, the narrower the peak and hence the transition between the constant and $x^{-2}$ behaviors is more abrupt). Initially, $y =$ constant and the density $\rho_0$ and pressure $p_0$ for the ground state are $\propto a^{-6}$.
 When $x \gg x_0$ and if $x \gg 1/|C_0|$ then $y = C_0^{-1} x^{-2}$.  The pressure and the density then have two terms 
 \begin{eqnarray}
 \rho_0 &=& \frac{H_{0r}^{3/2}}{4 a^6 m^{1/2} C_0}  +  \frac{m^{5/2} C_0}{2 H_{0r}^{3/2}} \\ \nonumber
 p_0 &=& \frac{H_{0r}^{3/2}}{4 a^6 m^{1/2} C_0}  -  \frac{m^{5/2} C_0}{2 H_{0r}^{3/2}},
 \end{eqnarray}
where the $a^{-6}$ term dominates at early times. 
The density transitions to a cosmological constant with $p_0 = -\rho_0$ when $a \approx H_{0r}^{1/2} C_0^{-1/3} 2^{-1/6} m^{-1/2}$. 
If $\rho_\Phi > \rho_{\rm rad}$ then $\rho_\Phi \propto a^{-6}$ \citep{arby}. However, in this regime 
the above derivation becomes invalid. 
\subsection{Matter Domination}
In the matter domination regime ($n=3/2$), Eq.\ (\ref{scalar}) becomes
\begin{equation}
\frac{d^2 \psi_k}{d\tau^2} + \left(k^2 -  \frac{H_{0m}^2}{2 a}  + a^2 m^2\right) \psi_k =0,  ~
\end{equation}
where $H = H_{0m} a^{-3/2}$ with $H_{0m} \approx 7.8 \times 10^{-34} eV$. 

Using Eq.\ (\ref{psibreak}) we obtain an equation for $\omega_k$:
 \begin{eqnarray}
 - \frac{\omega_k a}{2} \frac{d^2 \omega_k}{d a^2} - \frac{\omega_k}{4} \frac{d \omega_k}{da} +
 \frac{3 a}{4} \left(\frac{d\omega_k}{da}\right)^2 - \frac{\omega_k^4}{H_{0m}^2}  \\ \nonumber
 + \omega_k^2\left[\left(\frac{k}{H_{0m}}\right)^2 - \frac{1}{2a} + \left(\frac{a m}{H_{0m}} \right)^2 \right] =0.
 \end{eqnarray}
 For the ground state ($k=0$), this equation has an exact solution
 \begin{equation}
 \omega_0 = \frac{a m}{C_1 \sin[4 m a^{3/2}/(3 H_{0m}) + \alpha] + C_2},
 \label{matter_sol}
 \end{equation}
 where $C_1$, $C_2$ and the phase $\alpha$ are constants with the constraint $C_2^2 - C_1^2 = 1$. This solution is in agreement with \cite{arby}.
 When $C_1=0$ the solution reduces to $\omega_0 = am$. Solutions with non-zero $C_1$ oscillate around the $\omega_0 = am$ solution. Eq.\ (\ref{matter_sol})  can also be written in terms of $t$ as
 \begin{equation}
 \omega_0 = \frac{a m}{C_1 \sin(2 m t + {\rm \alpha}) + C_2}.
 \end{equation}
The oscillations have a period of $\pi / m$ ($\sim$ a few years for $m = 10^{-23}$ eV). The pressure averages to zero on cosmological timescales causing the ground state scalar field particles to behave like pressureless matter.

 \subsection{Numerical Solution}
  We also solve Eq.\ (\ref{fullw}) numerically including the effect of both radiation and matter in the Hubble parameter, 
 \begin{equation}
 H = \sqrt{H_{0r}^2 a^{-4} + H_{0 m}^2 a^{-3}} ~.
 \end{equation}
 The numerical solutions are fully specified by the value of the field and its first derivative at a given $a$, as well as an overall scaling of the density.  In both cases, the overall scaling of the density was chosen to match the observed cosmological density of cold dark matter.

Two representative solutions are shown in Fig.\ \ref{fig1}. The initial conditions in terms of the $x$ and $y$ variables introduced above are $y(10^{-8}) \approx 0.48$ and $y'(10^{-8}) = 0.49$ for solution 1, and $y(10^{-8}) \approx 5\times 10^3$ and $y'(10^{-8}) \approx 5. \times 10^7$ for solution 2.   Fig.\ \ref{fig1}(a) shows the ground state density as a function of scale factor, along with the radiation density $\rho_{\rm rad}$.  Fig.\ \ref{fig1}{b} displays $w = p_{DM}/ \rho_{DM}$. The density at early times decays like $a^{-6}$ and is determined by the mass of the scalar particle and the required density in the ground state at late times.  Solution 1 requires a large density at early times which is not compatible with the standard $\Lambda$CDM model.  Solution 2 has a phase where the ground state density is constant. This allows the initial density to be much lower and therefore compatible with the standard model up to the the approximate time of nucleosynthesis ($a \approx 10^{-10}$). The final density is the same, but $w$ oscillates at late times around the pressureless $w = 0$ solution. 
\section{Conclusion}
In this paper we assume dark matter is composed of ultralight scalar particles ($m=10^{-23}$ eV) with 
a Compton wavelength of galactic scales. Halos formed from such particles naturally do not 
exhibit small scale structure, avoiding the over-abundance of dwarf galaxies. 
\cite{urena-lopez} showed that for $m < 10^{-14}$ eV Bose-condensation always occurs. 
The condensate has the correct cosmological behavior today. When $H<m$, particles in the
ground state behave like presureless matter while particles in excited states act as radiation. 
When $H>m$, the dark matter density is initially $\rho_{\Phi} \propto a^{-6}$ 
and then switches to a cosmological-constant behavior \citep{arby}. The density of the excited 
states remains radiation-like $\rho_{\rm \Phi \; ex} \propto a^{-4}$ until 
$T \sim H$ ($a \sim 10^{-32}$) and is sub-dominant to the density of known radiation. 

We find that if these particles decouple from regular matter before Standard Model particles annihilate, 
their temperature today is $T_\Phi \approx 0.9$ K. This temperature is substantially lower than the temperature of the 
CMB and neutrinos, leaving nucleosynthesis unaffected.  It is consistent with cosmological constraints
on the amount of hot dark matter in the universe from WMAP 7 year+BAO+H0 observations, which yield a one-sigma
upper limit of $T_\Phi \lesssim 2.25$ K that is independent of particle mass. After decoupling the scalar field has no 
interactions and hence cannot be detected by particle physics experiments or precision tests of gravity. 

A challenging requirement for our model is the large particle-antiparticle asymmetry that may arise from 
statistical fluctuations in the early universe or from asymmetric reactions before decoupling. 
Understanding this better is the subject of future work. A more obvious challenge is the low mass, which
seems unnatural from a particle physics perspective. However, even lighter scalar fields ($m \sim 10^{-33} eV$) 
have been proposed to explain the accelerated expansion of the universe \citep{m33_1,m33_2}.

\section*{Acknowledgements}
 We thank Duncan Brown, Sam Finn, Ed Seidel, Ira Wasserman and Ed Witten for inspiring discussions. 
This research is partially funded by NSF grants Nos. PHY 06-53462 and PHY-0847611.

\end{document}